\documentstyle[aps,twocolumn,psfig,prl]{revtex}
\begin{document}
\draft
\wideabs{
\title{Possible explanation for star-crushing effect in 
binary neutron star simulations}  

\author{\'Eanna \'E.\ Flanagan$^{*}$}
\address{Newman Laboratory of Nuclear Studies, 
Cornell University, Ithaca, NY 14853-5001.}  

\maketitle


\begin{abstract}
A possible explanation is suggested for the controversial
star-crushing effect seen in numerical simulations of inspiraling
neutron star binaries by Wilson, Mathews and Marronetti (WMM).  
An apparently incorrect definition of momentum density in the momentum
constraint equation used by WMM gives rise to a post-1-Newtonian error
in the approximation scheme.  We show by means of an analytic,
post-1-Newtonian calculation that this error causes an increase of
the stars' central densities which is of the order of several percent
when the stars are separated by a few stellar radii, in agreement with
what is seen in the simulations.  
\end{abstract}

}

\def\beq{\begin{equation}}
\def\endeq{\end{equation}}



A controversial issue in the astrophysics community recently
has 
been the claim by Wilson, Mathews and Marronetti (WMM), based on
numerical simulations, that inspiraling binary neutron stars are subject to a
general-relativistic crushing force that cause them to individually
collapse to black holes before they merge
\cite{wmm1,wmm2,wmm3,wmm4}.  Such a crushing force, if it
existed, would have profound implications for current efforts to detect
gravitational waves from such systems with LIGO, VIRGO and other
ground based detectors.  The WMM claim has been disputed by several
researchers utilizing a variety of approximate analytical and
numerical techniques
\cite{lai,stu1,kip}, and
recent independent numerical simulations using the same approximation
scheme as WMM show no crushing effect \cite{bonn}.  In this paper we
suggest an explanation for the star-crushing effect perceived by WMM.

We start with the standard ADM equations.  The metric is
\beq
ds^2 = -(\alpha^2 - \beta_i\beta^i) dt^2 + 2 \beta_i dx^i dt +
\gamma_{ij}dx^i dx^j, 
\label{metric0}
\end{equation}
so that the lapse function is $\alpha$ and the shift vector is
$\beta^i$.  The extrinsic curvature $K_{ij}$ is given by
\begin{equation}
\dot \gamma_{ij} = -2\alpha K_{ij} + D_i \beta_j + D_j \beta_i,
\label{gdot}
\end{equation}
where $D_i$ is the derivative operator associated with $\gamma_{ij}$
and dots denote derivatives with respect to $t$.  The Hamiltonian
constraint is 
\begin{equation}
{}^{(3)}R - K_{ij}K^{ij} + K^2 = 16\pi \rho_H,
\label{ham}
\end{equation}
where ${}^{(3)}R$ is the Ricci scalar of $\gamma_{ij}$, $K \equiv
\gamma^{ij} K_{ij}$, $n^\alpha$ is the normal to the $t = $ const
surface given by $n_\beta = -\alpha (dt)_\beta$, and $\rho_H = 
T_{\alpha\beta} n^\alpha n^\beta$.  The
momentum constraint is
\begin{equation}
D_i(K^{ij} - \gamma^{ij}K) = 8 \pi S^j,
\label{evans37}
\end{equation}
where 
$
S^\alpha \equiv - h^{\alpha\beta} T_{\beta\gamma} n^\gamma
$
and $h^{\alpha\beta} \equiv g^{\alpha\beta} + n^\alpha n^\beta$
is the projection tensor.  [Here Greek indices run over 
$(0,1,2,3)$ and Roman indices over $(1,2,3)$.] 
Finally the trace of the space-space part of Einstein's equation is
\begin{eqnarray}
4 \pi \alpha S &=& {\dot K} + D^i D_i \alpha - \beta^l D_l K \nonumber
\\
\mbox{} && 
- {\alpha \over
4} \left[ {}^{(3)}R + 3 K^{ij} K_{ij} + K^2 \right],
\label{spacespace2}
\end{eqnarray}
where $S \equiv h^{\alpha\beta} T^{\alpha\beta}$.

The main elements of the WMM approximation scheme are as follows
\cite{wmm1,wmm2,wmm3,wmm4}: 
(i) They use the standard perfect fluid equations to evolve
the fluid in the background metric (\ref{metric0}).
The stress-energy tensor is
\beq
T_{\alpha\beta} = ({\bar \rho}+p) u_\alpha u_\beta + p
g_{\alpha\beta},
\label{Tabdef}
\endeq
where $u^\alpha$ is the 4-velocity, $p$ is the
pressure and ${\bar \rho}$ is the energy density. 
The equations of motion are $\nabla_\alpha T^{\alpha\beta} = 0$ and 
$\nabla_\alpha (n u^\alpha)=0$, where $n$ is the baryon number
density.  (ii) They work in a co-rotating
coordinate system of the form 
(\ref{metric0}), so that the large $r$ boundary condition on the shift
vector is $\beta_i(x^j) = \epsilon_{ijk} \Omega^j x^k + O(r^0)$, where
$\Omega^j$ is the orbital angular velocity.  (iii) They impose that the
spatial metric $\gamma_{ij}$ be conformally flat, 
$
\gamma_{ij} = \varphi^4 {\bar \gamma}_{ij},
$
where ${\bar \gamma}_{ij}$ is flat and time independent.
By decomposing the extrinsic curvature as $K_{ij} = A_{ij} + K
\gamma_{ij}/3$ where $A_{ij}$ is traceless, and combining with
(\ref{gdot}) and the conformal flatness condition one gets
\beq
K = {- 6 {\dot \varphi} \over \varphi \alpha} + {1 \over \alpha} D_i
\beta^i
\label{trKeqn}
\endeq
and
\beq
A_{ij} = {1 \over 2 \alpha} \left[ D_i \beta_j + D_j \beta_i - {2
\over 3} (D_k \beta^k) \gamma_{ij} \right].
\label{Aijeqn}
\endeq
Using the relations (\ref{trKeqn}) and (\ref{Aijeqn}), the Hamiltonian
constraint (\ref{ham}), the momentum constraint (\ref{evans37}) and
the dynamical equation (\ref{spacespace2}) can be written
schematically as 
\begin{eqnarray}
{\dot {\varphi}} &=& F_1[\varphi,\beta^i,\alpha],
\label{scheme1} \\
0 &=& F_2^i[\beta^j,\alpha,\varphi,{\dot \varphi}],
\label{scheme2} \\
\label{scheme3}
{\ddot \varphi} &=& F_3[\varphi,{\dot \varphi},\alpha,\beta^k],
\end{eqnarray}
for some functionals $F_1$, $F_2^i$ and $F_3$.  (iv) They use a
quasi-equilibrium approximation scheme which means 
that they substitute ${\dot \varphi} = {\ddot \varphi} =0$ into Eqs.\
(\ref{scheme1})--(\ref{scheme3}).  (v) They substitute the 
maximal slicing condition $K = 0$ into the resulting equations.
This yields a system of equations in which one can solve for $\alpha$,
$\varphi$, and $\beta^i$ at each instant from $T_{\alpha\beta}$.

We now turn to a description of the apparent error in the momentum
constraint equation used by WMM.  Consider the following two
inequivalent definitions of momentum density.  The first is 
${\bar S}^\alpha \equiv - h^{\alpha\beta} T_{\beta\gamma} n^\gamma,
$
which is just the quantity which appears in
the momentum constraint (\ref{evans37}).  Using the perfect fluid
stress energy tensor 
(\ref{Tabdef}) 
and the notations $W \equiv - n_\beta u^\beta =  \alpha u^t$ and
$\sigma \equiv {\bar \rho} + p$, it can be written as
\beq
{\bar S}^\alpha = W \sigma h^{\alpha\beta} u_\beta.
\label{last}
\endeq
The second definition is simply the expression
(\ref{last}) without the projection tensor:
\beq
{\hat S}^\alpha = W \sigma u^\alpha.
\label{lastt}
\endeq
WMM appear to confuse the two different quantities (\ref{last}) and
(\ref{lastt}).  They define only a 3-vector $S_i$;
this definition [Eq.\ (47) of Ref.\ \cite{wmm2}] is
compatible with both definitions (\ref{last}) and (\ref{lastt}), since
${\bar S}_i = {\hat S}_i$ (but ${\bar S}_t \ne {\hat S}_t$).  
However, the 4-vector $S_\alpha$ appears in some of their equations.
Their hydrodynamic equations are correct only if their $S_\alpha$ is
interpreted to be ${\hat S}_\alpha$, while their momentum constraint
is correct only if $S_{\alpha}$ is interpreted to be ${\bar
S}_\alpha$.

This confusion apparently gives rise to an error in their equation for
the shift vector.  WMM solve for the shift vector by combining
the relation (\ref{Aijeqn}) with the assumption $K=0$ and with the
momentum constraint (\ref{evans37}).  The resulting equation for the
shift vector is 
\FL
\beq
{\bar D}^2 \beta^i + {1 \over 3} {\bar D}^i ({\bar D}_k \beta^k) = 
{\bar \Lambda}^{ij} {\bar D}_j \ln (\alpha/
\varphi^6) + 16 \pi \alpha \varphi^4 {\bar S}^j,
\label{forbeta0}
\endeq 
where ${\bar D}_i$ is the derivative operator associated with the flat
metric ${\bar \gamma}_{ij}$, and
$
{\bar \Lambda}^{ij} \equiv {\bar D}^i \beta^j + {\bar D}^j \beta^i - 2
({\bar D}_k \beta^k) {\bar \gamma}^{ij}/3.
$
Equation (\ref{forbeta0}) agrees with WMM's corresponding Eq.\ (33) of
Ref.\ \cite{wmm2}.  However, WMM then rewrite their variable $S^j$ in
terms of $S_j = W \sigma u_j$.  For the correct variable $S^j = {\bar
S}^j$, we have ${\bar S}^j = \varphi^{-4} {\bar S}_j = \varphi^{-4} W
\sigma u_j$.  For the incorrect variable ${\hat S}^j$ we have instead
\beq
{\hat S}^j = W \sigma \left[ \varphi^{-4} u_j - {W \over \alpha}
\beta^j \right].
\label{wrongans}
\endeq
Inserting Eq.\ (\ref{wrongans}) into Eq.\ (\ref{forbeta0}), WMM obtain
the equation [Eq.\ (41) of Ref.\ \cite{wmm2}, also Eq.\ (15) of Ref.\
\cite{wmm4}]
\begin{eqnarray}
{\bar D}^2 \beta^j +&& {1 \over 3} {\bar D}^j ({\bar D}_k \beta^k) = 
{\bar \Lambda}^{ij} {\bar D}_i \ln (\alpha /
\varphi^6) \nonumber \\
\mbox{} && 
+ 16 \pi \alpha \varphi^4 W \sigma \left[ \varphi^{-4} u_j -
\epsilon_0 {W \over \alpha} \beta^j \right],
\label{forbeta1}
\end{eqnarray}
with $\epsilon_0 \equiv 1$ \cite{typonote}.  
The correct version of this equation is given by $\epsilon_0=0$; see,
for example, Eq.\ (2.16) of Ref.\ \cite{stu1}.

We now turn to calculating the leading order effect of this error on
the stars' central densities.  We define a {\it fictitious
stress-energy tensor} $\Delta T_{\alpha\beta}$ by
$
G_{\alpha\beta}[g_{\mu\nu}] = 8 \pi \left[ {}^{(F)} T_{\alpha\beta} + \Delta
T_{\alpha\beta}\right], 
$
where $g_{\mu\nu}$ is the metric obtained by solving the WMM
equations and ${}^{(F)} T_{\alpha\beta}$ is the fluid stress-energy
tensor (\ref{Tabdef}).  
It is a useful point of view to regard Einstein's equation as being
satisfied exactly, but with an extra type of matter present whose
(conserved) stress tensor is $\Delta T_{\alpha\beta}$ and which
interacts with the neutron stars only gravitationally.  
The fictitious stress-energy tensor is of post-1-Newtonian
order, although without the error term in Eq.\ (\ref{forbeta1}) it
would have been of post-2-Newtonian order.  Our approach will be to
calculate $\Delta T_{\alpha\beta}$ analytically to 
post-1-Newtonian order, and then, starting from a correct,
post-1-Newtonian description of the binary, to solve for the
perturbation to the stellar structure that is linear in $\Delta
T_{\alpha\beta}$.  

In our calculations, it will be sufficient to restrict attention to
stationary solutions for which the vector field $\partial /\partial t$
is a killing vector field, since the numerical, dynamic solutions to the
WMM equations relax to such stationary states \cite{wmm2}.
There are two contributions to $\Delta
T_{\alpha\beta}$: (i) a direct contribution due to the error term in
Eq.\ (\ref{forbeta1}), and (ii) an indirect contribution due to the
fact that the first error causes a non-zero $K$ and invalidates the
maximal slicing assumption.  

We first calculate the direct contribution.  We can decompose
the fictitious stress energy tensor as 
\FL
\beq
\Delta T^{\alpha\beta} = \Delta \rho \, n^\alpha n^\beta + 2 n^{(\alpha}
\Delta S^{\beta)} + \Delta S^{\alpha\beta}_{({\rm TF})} + {1 \over 3}
h^{\alpha\beta} \Delta S,
\label{stsplit}
\endeq
where
$\Delta \rho \equiv \Delta T_{\alpha\beta} n^\alpha n^\beta$, 
$\Delta S^\alpha \equiv 
- (g^{\alpha\beta} + n^\alpha n^\beta) \Delta T_{\beta\gamma}
n^\gamma$, $\Delta S \equiv (g^{\alpha\beta} + n^\alpha n^\beta) \Delta
T_{\alpha\beta}$ and $\Delta S^{\alpha\beta}_{(TF)}$ is orthogonal to
$n^\alpha$ and tracefree.  
We define $\Delta \beta^i \equiv \beta^i - 
{\bar \epsilon}^{ijk} \Omega_j x_k$, where ${\bar \epsilon}^{ijk}$ is
the volume element associated with the flat metric ${\bar
\gamma}_{ij}$ and $x^i$ are Cartesian coordinates associated with
${\bar \gamma}_{ij}$; thus $\Delta \beta^i \to 0$ as $r \to \infty$.
Rewriting Eq.\ (\ref{forbeta1}) in terms of $\Delta \beta^i$ and 
the contravariant components of the 4-velocity and taking the
post-1-Newtonian limit yields
\begin{eqnarray}
{\bar D}^2 \Delta \beta^j +&& {1 \over 3} {\bar D}^j ({\bar D}_k
\Delta \beta^k) =  \nonumber \\
\mbox{} &&
 16 \pi \rho_M \left[ v^j
+ (1 - \epsilon_0) ({\bf \Omega} \times {\bf x})^j \right],
\label{forbeta4}
\end{eqnarray}
where $\rho_M$ is the Newtonian mass density and $v^i = d x^i/dt$ is
the 3-velocity in the rotating frame (\ref{metric0}).  From Eq.\
(\ref{forbeta4}) we see that there is there is a direct contribution
\beq
- \epsilon_0 ({\bf \Omega} \times {\bf x} ) \rho_M
\label{direct}
\endeq
to the quantity $\Delta S^i$.  

Consider now the indirect contribution.  Using Eq.\ (\ref{trKeqn}) and
the stationarity condition ${\dot \varphi}=0$ yields
\beq
K = {1 \over \alpha} \left[ {\bar D}_i \beta^i + 6 \beta^i {\bar D}_i
\ln \varphi \right].
\label{Keqn1}
\endeq
We now solve Eq.\ (\ref{forbeta4}) for the quantity ${\bar D}_i
\beta^i = {\bar D}_i \Delta \beta^i$, insert the result into Eq.\
(\ref{Keqn1}), make use of the Newtonian continuity equation in the
rotating frame ${\bar D}_i (\rho_M v^i) = - {\dot \rho_M} =0$, and use
the post-1-Newtonian relation $\varphi = 1 - \Phi/2$ between the
conformal factor $\varphi$ and the Newtonian potential $\Phi$.  The
result is
\beq
K = - 3 \epsilon_0 ( {\bf \Omega} \times {\bf x} ) \cdot {\bf \nabla}
\Phi.
\label{Kans}
\endeq
Now WMM insert the assumption $K=0$ into 
Eqs.\ (\ref{ham}), (\ref{evans37}) and (\ref{spacespace2}).
Since $K$ is actually
non-vanishing, this gives rise to the following contributions to
$\Delta T_{\alpha\beta}$: $\Delta \rho = K^2 / 24 \pi$, $\Delta S^i =
- D^i K / 12 \pi$, and $\Delta S = ({\dot K} - \beta^i D_i K - \alpha
K^2 /2)/4 \pi \alpha$.  Using the relation (\ref{Kans}), taking the
post-1-Newtonian limit, adding the direct contribution (\ref{direct}),
using the stationarity assumption and letting $\epsilon_0 \to 1$
finally yields \cite{offdiagonal} $\Delta \rho  = 0$,
\begin{eqnarray}
%
\label{fictitiousST0ia}
\Delta {\bf S} &=& {1 \over 4 \pi} {\bf \nabla} \cdot \left[ ( {\bf \Omega}
\times {\bf x} ) \cdot {\bf \nabla} \Phi \right]
- ({\bf \Omega} \times {\bf x} ) \rho_M, \\
\label{fictitiousSTiia}
\Delta S &=& {3 \over 4 \pi} \left[ \left( {\bf \Omega} \times {\bf x}
\right) \cdot {\bf \nabla} \right]^2 \Phi.
\end{eqnarray}

We next calculate the effect of the fictitious stress energy tensor
(\ref{fictitiousST0ia})--(\ref{fictitiousSTiia}) on the neutron
stars' central densities.  Focus attention on one of the two stars,
say star A.  We define the two
dimensionless parameters $\epsilon \equiv M / R$ and $\kappa \equiv R
/ L$, where  $M$ is the mass and $R$ the radius of either star, and
$L$ is the orbital separation.  We will work to the leading
non-vanishing order in $\kappa$, which will turn out to be linear in
$\kappa$.  Now it is known that the leading order
(tidal) fractional corrections to the internal structure of star A due
to the other star scale as $\kappa^3$, to post-1-Newtonian order as
well as in 
Newtonian gravity \cite{kip}; we can neglect these corrections.
Hence, accurate to $O(\kappa^2)$, we can find a
non-rotating coordinate system $({\bar t},{\bar x}^i)$ near star A in
which the metric is that of an isolated neutron star.  
These coordinates are related to the original co-rotating
coordinates $(t,x^i)$ of the line element (\ref{metric0}) by
\begin{eqnarray}
\label{coordtransf1}
t &=& {\bar t} + {\dot z}^i({\bar t}) {\bar x}^i \\
R_{ij}(t) x^j &=& z^i({\bar t}) + {\bar x}^i.
\label{coordtransf2}
\end{eqnarray}
Here $R_{ij}(t)$ is the rotation matrix satisfying ${\dot R}_{il}(t) =
\epsilon_{ijk} \Omega_j R_{kl}(t)$, $z^i(t) \equiv R_{ij}(t) z^j_A$,
and $z^j_A$ is the (time-independent) coordinate location of the
center of star A in the $(t,x^i)$ coordinates.  The transformation
(\ref{coordtransf1})--(\ref{coordtransf2}) is approximate but is
sufficiently accurate for our calculation.

Next, we combine Eqs.\ (\ref{stsplit}) and
(\ref{fictitiousST0ia})--(\ref{fictitiousSTiia}) together with
$n^\mu=(1,-\beta^i)/\alpha\approx(1,-{\bf \Omega}\times {\bf x})$
to obtain the contravariant components of $\Delta T^{\alpha\beta}$ in
the $(t,x^i)$ coordinate 
system.  We then use the zeroth order metric $ds^2 = - dt^2 +
2 ({\bf \Omega} \times {\bf x})_i \,dt \,dx^i + d {\bf x}^2$ to obtain the
covariant components $\Delta T_{\alpha\beta}$, and finally use the
transformation (\ref{coordtransf1})--(\ref{coordtransf2}) to calculate
the components $\Delta T_{{\bar \alpha}{\bar \beta}}$ in the $({\bar
t},{\bar x}^i)$ coordinates, discarding all post-2-Newtonian terms.
The result is $\Delta T_{{\bar t}{\bar   
t}} = \Delta \rho_F$, $\Delta T_{{\bar t}{\bar i}} = - \Delta S^i$ and
$\Delta T_{{\bar i}{\bar j}} = \Delta p_F \delta_{{\bar i}{\bar j}} +
Q_{{\bar i}{\bar j}}$,
where $\Delta S^i$ is given by Eq.\ (\ref{fictitiousST0ia}), 
\beq
\Delta \rho_F = - {1 \over 2 \pi} \left[ \left( {\bf \Omega} \times
{\bf x} \right)   \cdot {\bf \nabla} \right]^2 \Phi
+ 2 ({\bf \Omega} \times {\bf x})^2 \rho_M
\label{fd}
\endeq
is the fictitious density,
\beq
\Delta p_F =  {1 \over 12 \pi} \left[ \left( {\bf \Omega} \times
{\bf x} \right)   \cdot {\bf \nabla} \right]^2 \Phi
+ {2 \over 3} ({\bf \Omega} \times {\bf x})^2 \rho_M
\label{fp}
\endeq
is the fictitious pressure, and where $Q_{{\bar
i}{\bar j}}$ is a traceless tensor that does not contribute to the
leading order change in central density.  In deriving Eqs.\
(\ref{fd})--(\ref{fp}) [but not in deriving the expression
(\ref{fictitiousST0ia}) for $\Delta T_{{\bar t}{\bar i}}$]
we replaced ${\bf \Omega} \times {\bf z}_A$ by
${\bf \Omega} \times {\bf x}$, which is valid to leading order in
$\kappa$.

Consider now the case where star A is non-rotating and hence
spherically symmetric.  Then, the fictitious momentum density $\Delta
T_{{\bar t}{\bar i}}$ will not affect the central density of the star
\cite{notewhy}, so we can restrict attention to $\Delta \rho_F$ and
$\Delta p_F$.  To leading order in $\kappa$, we can replace
$\Phi$ and $\rho_M$ in Eqs.\
(\ref{fd})--(\ref{fp}) by the self potential $\Phi_A$ and
the mass density $\rho_{M,A}$ of star A, 
and we can replace the quantity
${\bf \Omega} \times {\bf x}$ by ${\bf \Omega} \times {\bf z}_A$.
The first term in Eqs.\ (\ref{fd}) is then proportional to
$\partial^2 \Phi_A / \partial y^2$ with a suitable choice of $y$-axis,
which becomes ${\bf \nabla}^2 \Phi_A / 3 = 4 \pi \rho_{M,A}/3$ when we
average over solid angles about the center of star A.  Hence to
leading order in $\kappa$ we obtain
\begin{eqnarray}
\label{fd1}
\Delta \rho_F &=& {4 \over 3} \ v_{\rm orb}^2 \ \rho_{M,A} ,\ \ \ \ 
\Delta p_F = {7 \over 9} \ v_{\rm orb}^2 \ \rho_{M,A},
\label{fp1}
\end{eqnarray}
where ${\bf v}_{\rm orb} = {\bf \Omega} \times {\bf z}_A$ is the
orbital velocity.

\begin{figure}
{\psfig{file=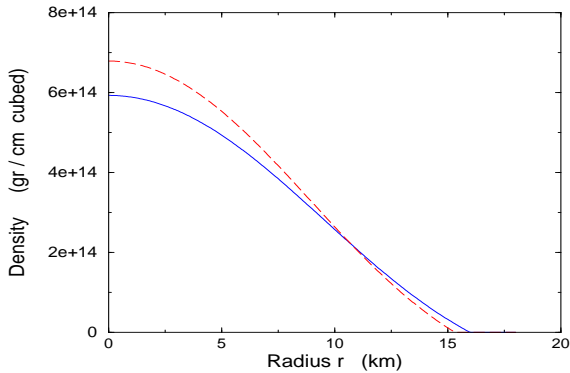,height=5cm,width=7.5cm,angle=0}}
\vskip 0.5cm
\caption{
Consider a neutron star described by a polytropic $\Gamma=2$ equation
of state, at an orbital separation of 4 stellar radii from another
similar neutron star.  The density profile of such a star will be very
close to that of an isolated star, which 
is shown as the solid line.  When the leading order
effects of the erroneous term in the momentum constraint equation are
included, the result is the dashed line.
}
\label{fig:profile}
\end{figure}

It is clear that the fictitious density and pressure
(\ref{fd1}) will cause a fractional increase in the
central density of star A proportional to $\epsilon \, \kappa$ at leading
order.  To evaluate the constant of proportionality we solve for the
perturbation to the structure of star A 
using the following modified form of the TOV equations \cite{hh}:
\begin{eqnarray}
\label{TOV1}
{d m \over d r} &=& 4 \pi r^2 \left[ \rho + \Delta \rho_F
\right],\\
\label{TOV2}
{d p \over d r} &=& - {(\rho + p) \left[
m + 4 \pi r^3 (p + \Delta p_F
) \right]\over r(r - 2 m) }.
\end{eqnarray}
Our procedure consists of: (i) solving 
Eqs.\ (\ref{TOV1})--(\ref{TOV2}) without the correction terms to obtain
the unperturbed structure of the star.  We use the same stellar model
as used in Ref.\ \cite{wmm4}, described by the polytropic equation
of state $p = K \rho_M^\Gamma$, $\rho = \rho_M + K \rho_M^\Gamma /
(\Gamma-1)$, where $\rho_M$ is the rest-mass density, $\Gamma =2$, and
$K = 1.8 \times 10^5$ erg cm$^3$ gr$^{-2}$.  We choose 
a central density for the unperturbed star of
$\rho_c = 5.93 \times 10^{14}$ gr
cm$^{-3}$, which implies a baryonic mass of $1.62 M_\odot$ and a total
mass of $1.51 M_\odot$.  (ii) We use Eqs.\ (\ref{fd1}) to
calculate $\Delta \rho_F$ and $\Delta p_F$.  (iii) We insert these
$\Delta \rho_F$ and $\Delta p_F$ into into Eqs.\
(\ref{TOV1})--(\ref{TOV2}), and adjust the choice of central density
$\rho_c$ until a perturbed stellar model with the same total baryonic
mass of $1.62 M_\odot$ is obtained.  The result is shown in Fig.\
\ref{fig:profile}, where we choose the stellar separation to be given
by $\kappa = R/L = 1/4$, corresponding to $v_{\rm orb} = 0.132$.  The
central density has increased by $\sim 15\%$.

{\vskip 0.3cm}
{\vskip 0.25cm}

Turn next to the case when star A is rigidly co-rotating.  The
fractional corrections to its internal structure due to its own
rotation scale as $\kappa^3$, and therefore to leading order in
$\kappa$ the above analysis of the effects of $\Delta p_F$ and $\Delta
\rho_F$ is still valid.  However, there is now in addition a
gravitomagnetic interaction between the fluid's velocity and the
fictitious momentum density $\Delta T_{{\bar t}{\bar i}}$.  A
straightforward computation shows that the radial component of the
gravitomagnetic force averaged over solid angles is
\beq
- {8 \over 3} \rho_M \Omega^2 r \left[ 2 \lambda(r) + r \Phi'(r)
  \right],
\endeq
where $\lambda(r)$ is given by $(r \partial_r + 3) \lambda = \Phi$
with $\lambda$ finite as $r \to 0$.  This force gives rise to a fractional
change in central density proportional to $\epsilon \, \kappa^2$.  
Evaluating this change numerically at $\kappa=1/4$ for the same
stellar model as above gives a contribution to $\delta \rho_c/\rho_c$
of less than one percent.  Therefore, the dominant contribution to
$\delta \rho_c$ should be that from $\Delta \rho_F$ and $\Delta p_F$,
and the crushing effect should be seen in the co-rotating case as well
as in the non-rotating case.

To conclude, we compare our predictions with the
behavior seen in the WMM simulations:  (i) The predicted magnitude of
$\delta \rho_c/\rho_c$ agrees with that seen.  (ii) The scaling $\delta
\rho_c \propto \kappa \propto 1/L$ is not inconsistent with the
scaling seen in the simulations \cite{notescaling}.  (iii) Our
analysis cannot explain the claim by WMM \cite{wmm3} that the crushing
effect is not seen in the co-rotating case.  In any case, it should be
straightforward to verify or falsify our proposed explanation by
re-running the simulations without the extra term in the momentum
constraint equation.

I thank Greg Cook, Grant Mathews, Saul Teukolsky and Ira Wasserman for
helpful discussions.  This research was supported in part by NSF grant
PHY--9722189 and by a Sloan Foundation fellowship.


\begin{references}
\bibitem{typonote}
Equations (41) of Ref.\ \cite{wmm2} and (15) of Ref.\
\cite{wmm4} actually differ from Eq.\ (\ref{forbeta1}) by factors of
$\varphi^4$ and $\varphi^{-4}$ in the first and second terms in the
square brackets.  These factors are possibly typos; in any case their
presence or absence changes the system of equations only at
post-2-Newtonian order and hence is irrelevant for the discussion of
this paper.
\bibitem{offdiagonal}
There is in addition a post-1-Newtonian contribution to the quantity
$\Delta S_{\alpha\beta}^{(TF)}$ which we ignore, since it can be shown
that this contribution does not affect the stars' central densities
at leading order.
\bibitem{notewhy} The gravitomagnetic interaction vanishes since the
fluid velocity is zero.  There is a force term proportional to the
time derivative of the transverse part of the perturbation to
$g_{{\bar t}{\bar i}}$, but one can show that the radial component of
this force has no spherically symmetric component.
\bibitem{hh}
It would be more consistent to use the post-1-Newtonian version of the
TOV equations rather than the fully relativistic version, but the
results are insensitive to which version we pick.
\bibitem{notescaling}
Although Ref.\ \cite{wmm4} suggests that $\delta \rho_c \propto
\kappa^2$, Fig.\ 2 of Ref.\ \cite{wmm4} is not incompatible with a fit of
the form $\delta \rho_c / \rho_c = \alpha_0 \kappa + \alpha_1
\kappa^2 + O(\kappa^3)$ with the two terms comparable at $\kappa =
1/4$.  This paper calculates only the leading order term $\alpha_0
\kappa$.
\end{references}
\end{document}